\documentclass[]{spie} 
 
\usepackage{amsmath,amsfonts,amssymb}
\usepackage{graphicx}
\usepackage[colorlinks=true, allcolors=blue]{hyperref}

\title{Improving VLT/SPHERE without additional hardware: Comparing quasi-static correction strategies.}

\author[a]{Axel Potier}
\author[b]{Zahed Wahhaj}
\author[c]{Raphael Galicher}
\author[c]{Johan Mazoyer}
\author[c]{Pierre Baudoz}
\author[d]{Gael Chauvin}
\author[a]{Garreth Ruane}

\affil[a]{Jet Propulsion Laboratory, California Institute of Technology, 4800 Oak Grove Dr., Pasadena, CA 91109, USA}
\affil[b]{European Southern Observatory, Alonso de Córdova 3107, Vitacura, Casilla 19001, Santiago, Chile}
\affil[c]{LESIA, Observatoire de Paris, CNRS, Université Paris Diderot, Université Pierre et Marie Curie, 5 place Jules Janssen, 92190 Meudon, France}
\affil[d]{Laboratoire Lagrange, Université Cote d’Azur, CNRS, Observatoire de la Cote d’Azur, 06304 Nice, France}

\authorinfo{Send correspondence to Axel Potier: \href{mailto:axel.q.potier@jpl.nasa.gov}{axel.q.potier@jpl.nasa.gov}\\ © 2022. All rights reserved. }

\begin{document} 
\maketitle

\begin{abstract}
Direct imaging is the primary technique currently used to detect young and warm exoplanets and understand their formation scenarios. The extreme flux ratio between an exoplanet and its host star requires the use of coronagraphs to attenuate the starlight and create high contrast images. However, their performance is limited by wavefront aberrations that cause stellar photons to leak through the coronagraph and on to the science detector preventing the observation of fainter extrasolar companions. The VLT/SPHERE instrument takes advantage of its efficient adaptive optics system to minimize dynamical aberrations to improve the image contrast. In good seeing conditions, the performance is limited by quasi-static aberrations caused by slowly varying aberrations and manufacturing defects in the optical components. The mitigation of these aberrations requires additional wavefront sensing and control algorithms to enhance the contrast performance of SPHERE. Dark hole algorithms initially developed for space-based application and recently performed on SPHERE calibration unit have shown significant improvement in contrast. This work presents a status update of dark hole algorithms applied on SPHERE and the results obtained during the on-sky tests performed on February 15th 2022.
\end{abstract}


\section{Introduction}
\label{sec:intro}
Direct imaging of exoplanet is currently performed with large ground-based telescopes equipped with state-of-the-art coronagraph instruments\cite{Jovanovic2015,Macintosh2015a,Beuzit2019}\,. These instruments provide diffraction-limited point spread functions thanks to their extreme adaptive optics systems (XAO) while the coronagraph aims to attenuate the starlight to reveal the faint companions and/or circumstellar disks. Their capabilities are however limited by post-XAO wavefront residuals that enable starlight leakage through the coronagraph, hence creating bright stellar speckles on the science detector. Advanced post-processing techniques like angular (ADI)\cite{Marois2006}\,, spectral (SDI)\cite{Racine1999}\,, polarimetric (PDI)\cite{Kuhn2001}\,, and reference-star (RDI)\cite{Lafreniere2009} differential imaging are therefore used on raw images to further enhance detection capabilities by mitigating speckle noise in the images. These techniques are particularly efficient to remove speckles originating from instabilities and wavefront errors internal to the instruments like the non-common-path aberrations (NCPAs). But the broadly used ADI technique requires 1-2~h sequences of observations and suffers for self-subtraction of astrophysical sources at small angular separations. Mitigating static and quasi-static speckles using focal plane wavefront sensors is now an important study of research \cite{Martinache2014, Bottom2017, Galicher2019, Marois2020}\,. In this work, we present the first on-sky correction of the NCPAs on SPHERE with pair-wise probing (PWP) and electric field conjugation (EFC), originally developed for high-contrast imaging in stable environments like space. In Sec.~\ref{sec:algos}, we describe the algorithms used throughout this study and how they were implemented in the instrument. In Sec.~\ref{sec:internal_source}, we demonstrate the algorithms on the internal source to investigate the contrast limitation and apply those solutions while observing on-sky. In Sec.~\ref{sec:onsky}, we demonstrate the correction loop on-sky in a half DH. In Sec.~\ref{sec:cdi}, we use PWP alone to calibrate the static stellar speckles in post-processing. Finally in Sec.~\ref{sec:conclusion}, we discuss the combination of the different strategies and best practices to optimize the speckle calibration in accordance with the science objectives.

\section{Method and algorithms}
\label{sec:algos}

PWP and EFC algorithms have been extensively described in the literature\cite{Borde2006,GiveOn2007SPIE,Potier2020a}\,. Their combination in closed-loop aims to minimize the speckle intensity a region of the science image hence called the dark hole (DH). First, PWP probing estimates the focal plane electric field (E-field) through a temporal modulation of the speckle field, equivalently to conventional phase diversity techniques. PWP requires a high-order deformable mirror (DM) to introduce optical path differences (OPDs) called probes that coherently interfere with the speckle field we wish to correct. In the absence of a specific DM for this purpose, the OPDs are introduced while AO runs in closed-loop by modifying the Shack-Hartmann (SH) reference slopes, assuming the sensor remains linear. Such an assumption might be invalid in the case of a pyramid wavefront sensor planned for the upgrade of both VLT/SPHERE\cite{Boccaletti2020} and Gemini/GPI\cite{Chilcote2018}\,. The probes need to be chosen such that they create a different electric field at every location of the speckle field. As explained in a previous work\cite{Potier2020b}\,, the first probe is here the poke of one individual actuator, whose phase-shift is mostly unaffected by the coronagraph masks. The second probe is then the individual poke of a second actuator in the direct neighborhood of the first one. The second actuator is chosen depending on the correction zone as explained in the next paragraph. The algorithm requires each probe to be pushed and pulled with an image recorded via the science detector for each DM shape. PWP therefore requires 4 images per E-field estimation (or per iteration when used in closed-loop with EFC).

EFC is the E-field counterpart of phase conjugation in AO. Knowing the E-field with any focal plane wavefront sensor (e.g. either the Self-Coherent Camera\cite{Baudoz2006} or PWP can be used), we aim to create destructive interference in regions of the focal plane by applying the opposite of the estimated E-field, applying OPDs using the DM in the pupil plane. The technique is powerful when applied in half the field of view (called half DH or HDH) since it enables the correction of speckles originating from phase and amplitude aberrations as well as the residual diffraction pattern from the coronagraph masks\cite{Potier2020b}\,. The algorithm inverts a Jacobian that describes the effect of each actuator movement on the E-field in the science detector plane. Here, the Jacobian is model-based using Fourier optics and a simplified compact model of the system. The DH regions are chosen in agreement with the set of probes and depend on the science case: two actuators spread vertically (resp. horizontally) are chosen for top and bottom (resp. left and right) DH corrections. The presented DH here ranges from 100~mas to 650~mas from the optical axis. The inversion of the Jacobian is regularized with a smooth filter (Tikhonov regularization\cite{Pueyo2009}) whose knee is defined at the 400th singular value (when sorted in descending order). We use a global loop control gain of 0.2 to ensure convergence.

Focal plane wavefront sensing is performed in the H3 band ($\lambda_0 = 1667$~nm, $\Delta\lambda = 54$~nm) with IRDIS\cite{Dohlen2008} behind an Apodized Pupil Lyot Coronagraph\cite{soummer2005} (APLC) whose diameter of the central focal plane mask obstruction is 185~mas. On-sky experiments were performed during the second half of the night of February 15th 2022, starting at 5~a.m. UTC. The seeing averaged $\sim$0.8~arcsec at 500~nm through the night while the atmospheric coherence time remained around 10~ms. These good conditions enabled the use of the small pinhole for spatial filtering in front on the SH WFS \cite{Poyneer2004}. Images with 64~s of exposure time are all acquired on HIP~57013 (RA=11 41 19.79 , $\delta$= -43 05 44.40 , R=5.5, H=5.5).

\section{Strategy 1: NCPA calibration using the internal source unit}
\label{sec:internal_source}

\begin{figure}[t]
    \centering
    \includegraphics[width=\linewidth]{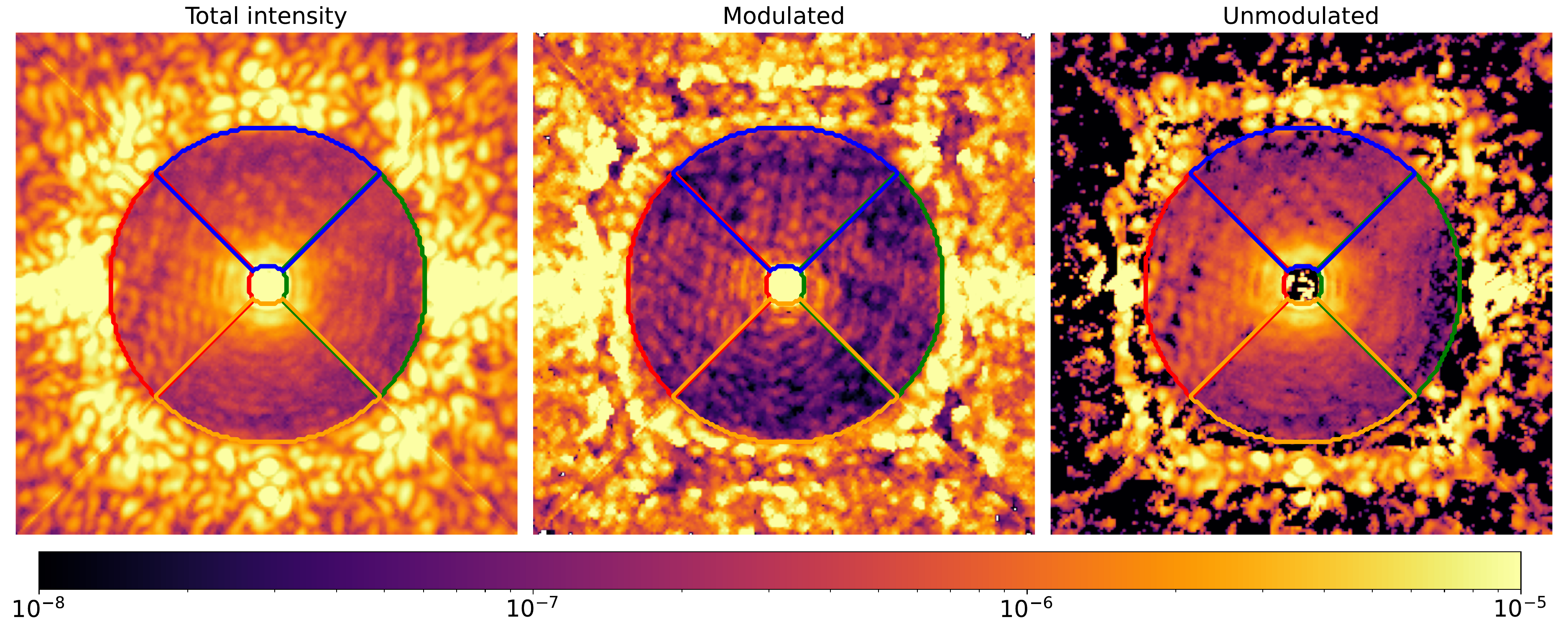}
    \caption{Internal calibration unit data. Left: Resulting image after half DH correction successively performed in the 4 quadrants of IRDIS field of view. Center: Remaining coherent speckle (also called modulated component) in the left image as sensed by PWP. Right: Residuals not sensed by PWP (Total intensity - Modulated component). The right image highlights the internal turbulent aberrations that cannot be sensed by PWP since the phase errors evolve too fast.}
    \label{fig:internalsouce_turbu}%
\end{figure}

\begin{figure}[t]
    \centering
    \includegraphics[width=9cm]{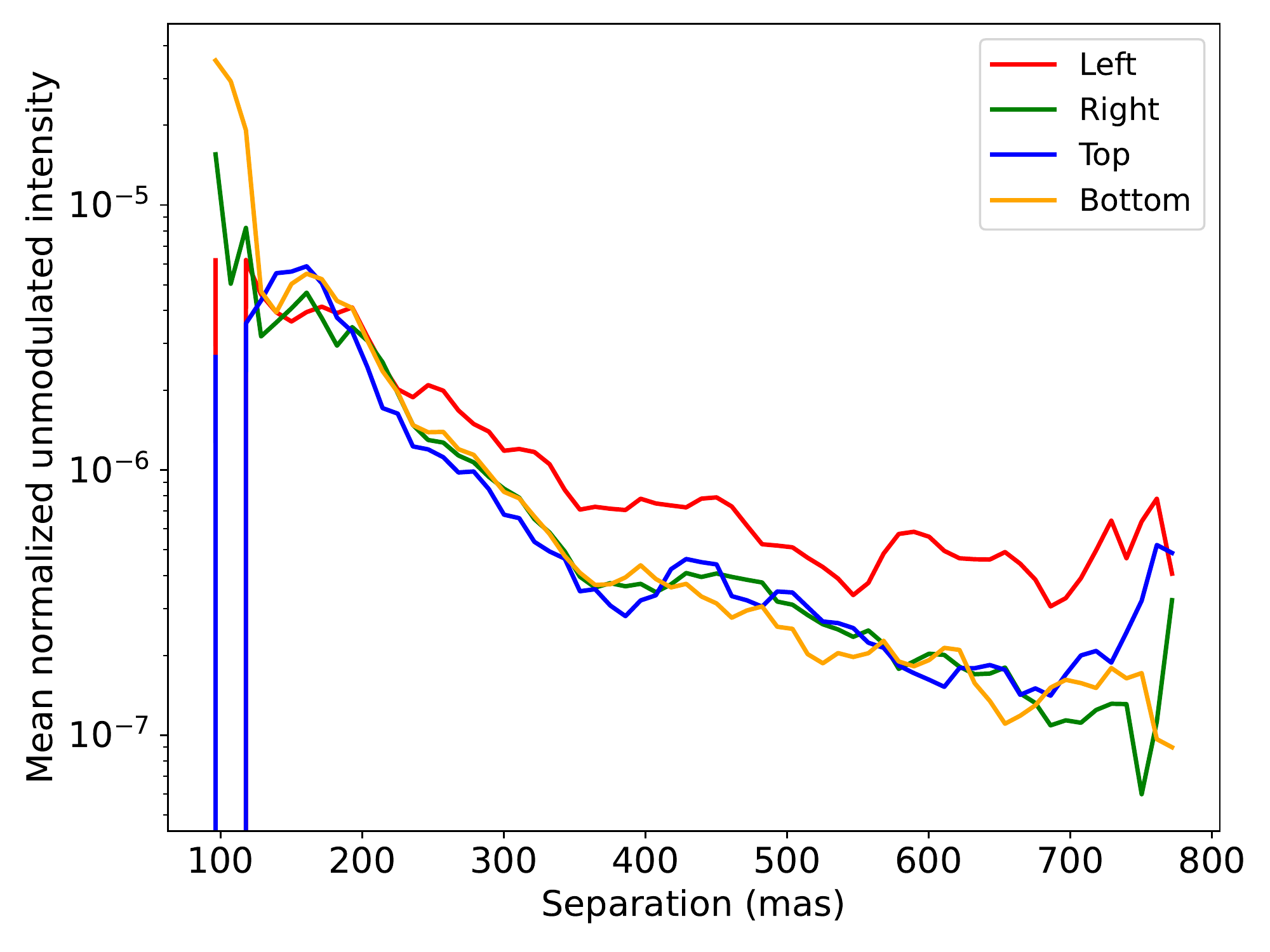}
    \caption{Internal calibration unit data. Normalized intensity of the unmodulated component (mainly the halo of internal turbulence) in the four image quadrants with respect to the angular separation from the optical axis.}
    \label{fig:internalsouce_contrastinco}%
\end{figure}

\subsection{Internal turbulence and ultimate contrast}
\label{subsec:internal_turbu}
Building on previous demonstrations using the internal source\cite{Potier2020b}, here we use a longer exposure time to improve on the estimate of the contrast floor. Our updated algorithm now enables the user to modify the science and probe image exposure times on the fly, which allowed us to dig deeper DH and better understand SPHERE limitations. We show in Fig.~\ref{fig:internalsouce_turbu} the composite image (10s exposure time) after 4 individual series of half DH corrections in 4 quadrants of the field of view. The images are normalized by the maximum of the non-coronagraphic point spread function (PSF). Unlike the expectations to reach performance down to $\sim10^{-8}$ as in other in-air testbeds in stabilized environments\cite{LlopSayson2020,Potier2020a,Baxter2021,Laginja2022}\,, the normalized intensity is limited by a smooth axi-symmetric halo whose level is $\sim10^{-6}$ at 200~mas. We also show the modulated component, i.e. the coherent part of the residuals that is sensed by PWP. It demonstrates the limitation is not caused by EFC or the inability of DM to correct for aberrations since the halo is not sensed by PWP. Some residual modulated speckles remain in each image quadrant and would be mitigated with additional iterations of PWP+EFC. The unmodulated component shown in Fig.~\ref{fig:internalsouce_turbu} represents the difference between the total intensity and the modulated component to highlight residuals that are not sensed by PWP. The unmodulated light is mostly a smooth halo whose intensity levels in the four different quadrants are shown in Fig.~\ref{fig:internalsouce_contrastinco}. The halo is explained as the effect of fast internal turbulence in VLT/SPHERE at a level of a few nanometers rms that has been recently discovered using the ZELDA wavefront sensor, and likely caused by a warm motor located underneath the NIR channel\cite{Vigan2022}\,. The limit of performance foreseen by the former study (from $10^{-5}$ to $10^{-7}$ in the HODM influence function) is here confirmed by the experiment. The difference of unmodulated intensity between the left quadrant and the three others could be explained by ghosts all oriented in the same direction during instrument commissioning.

\begin{figure}[t]
    \centering
    \includegraphics[width=12cm]{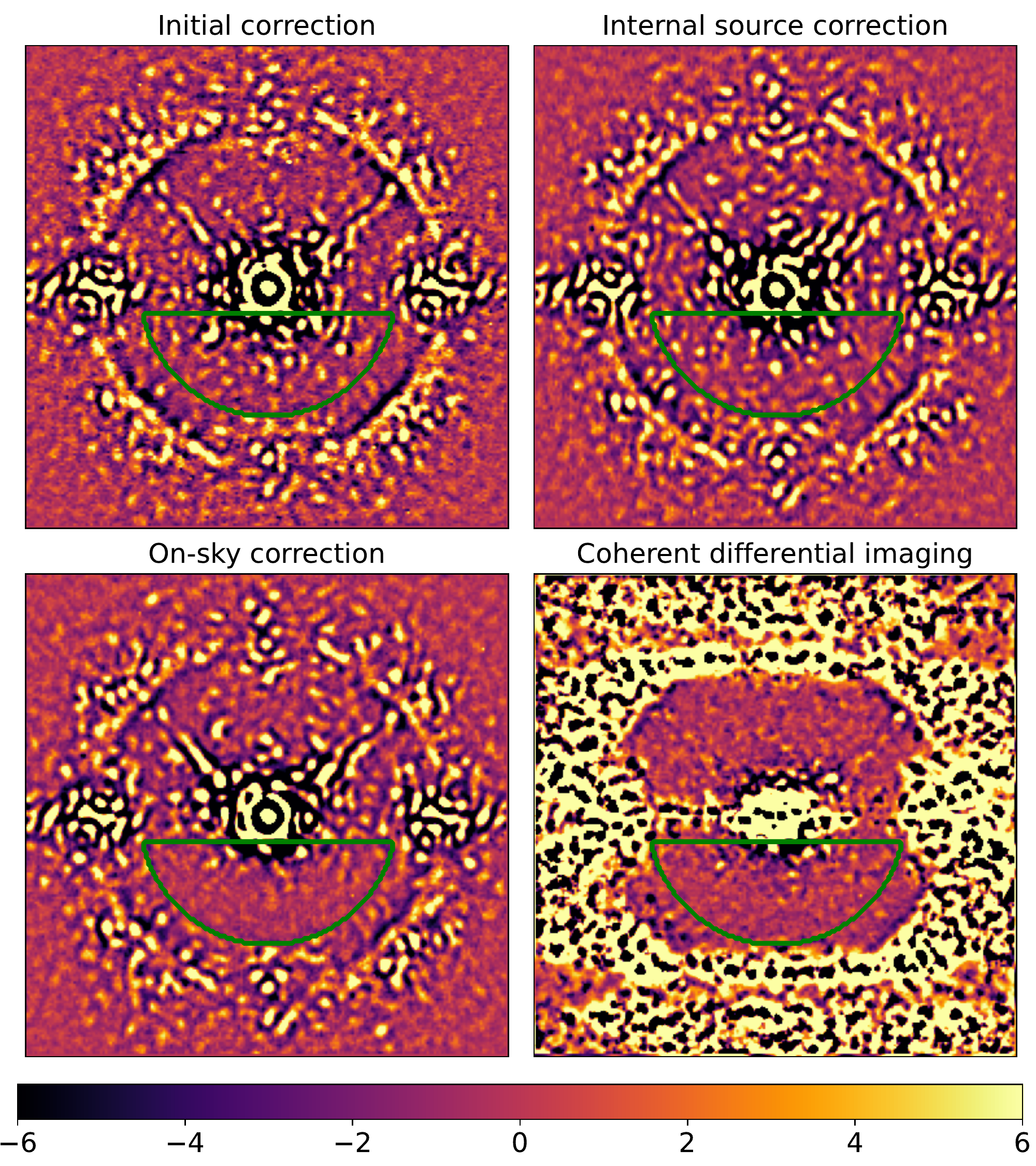}
    \caption{On-sky data ($\times10^{-6}$) with 64s of exposure time. Images before (top left) and after bottom half DH correction calculated either on the internal source (top right) or with the PWP+EFC loop closed on-sky (bottom left). The image on the bottom right is the result of coherent differential imaging after only one on-sky iteration of PWP+EFC. The halo of atmospheric residuals have been filtered in all the images.}
    \label{fig:comparison}%
\end{figure}

\subsection{On-sky performance applying the internal source correction}
\label{subsec:internal_onsky}
Our next test was to apply the DM settings that were determined \textit{a priori} using the internal source unit of SPHERE. Indeed, these settings do not require any telescope or instrument overheads during the night and can be acquired prior to any observation. The WFS slopes were recorded during the afternoon of February 15th, 2022 successively in the four image quadrants with a similar process as described in Sec.~\ref{subsec:internal_turbu}. The final WFS slopes were then recorded for each DH geometry and reapplied on sky a few hours later during the night observations. In Fig.~\ref{fig:comparison}, we compare the IRDIS images with the initial NCPA calibration and with the bottom DH correction calculated earlier on the internal source. The presented results are processed with a Gaussian high-passed filter whose standard deviation is equal to 0.57$\lambda/D$ in order to highlight the effect of localized static or quasi-static speckles in the images by reducing the effect of the smooth atmospheric turbulence halo.

\begin{figure}[t]
    \centering
    \includegraphics[width=8cm]{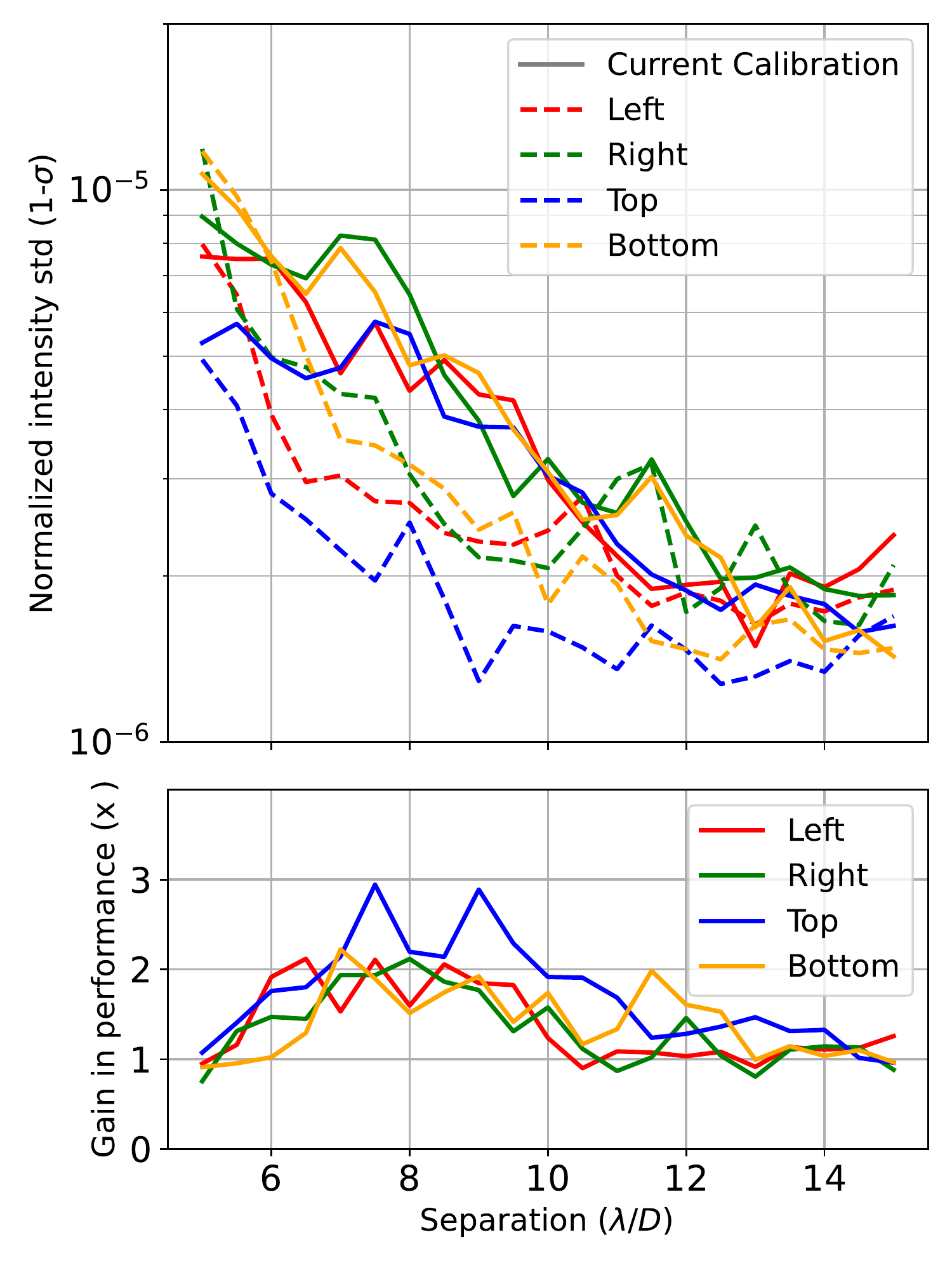}
    \caption{On-sky data. Top: 1-$\sigma$ standard deviation of the normalized intensity obtained before and after DH correction of the NCPAs in the four image quadrants versus the distance from the star. Bottom: Gain in performance with respect to the current SPHERE calibration by using the DH calibration calculated on the internal source unit.}
    \label{fig:internalsouce_onsky}%
\end{figure}

Qualitatively, the corrected image shows a slight decrease of the static speckle intensity in the DH region. We especially correct for the speckles induced by the diffraction pattern of the APLC coronagraph associated with large telescope spiders that can be considered as amplitude aberrations. Figure~\ref{fig:internalsouce_onsky} shows the image radial profiles RMS (calculated as the standard deviation in azimutal rings of size $\lambda/2D$ in the DH regions versus the angular separation) for corrections performed in the 4 image quadrants. The curves show that the best gain in performance is obtained with the top DH where an improvement up to a factor of 3 is attained in this region. In the other regions, performance improvement are limited up to a factor of 2 because the calibration on the internal source unit leads to 1) the insensitivity to quasi-static aberrations whose spatial distribution have evolved in between the calibration calculated on the internal source and the application of this calibration on-sky \cite{Vigan2022}\,, 2) a non-common-path between both the internal calibration unit and the telescope pupil, including a misalignment on the coronagraph focal plane mask as well as the insensitivity to amplitude aberrations located in the telescope light path, all resulting in the creation of new speckles during the on-sky observations, and 3) a discrepancy in the correction of the APLC diffraction pattern due to the inconsistency of entrance pupil between internal source and on-sky observations. First limitation could be partly mitigated by calculated the correction reference slopes right before the on-sky operations. The third limitation could be solved with the introduction of a VLT-pupil-like mask in the optical path of the internal calibration unit. The second limitation can only be resolved by closing the correction loop on sky.

\section{Strategy 2: on-sky closed-loop}
\label{sec:onsky}
\begin{figure}[t]
    \centering
    \includegraphics[width=8cm]{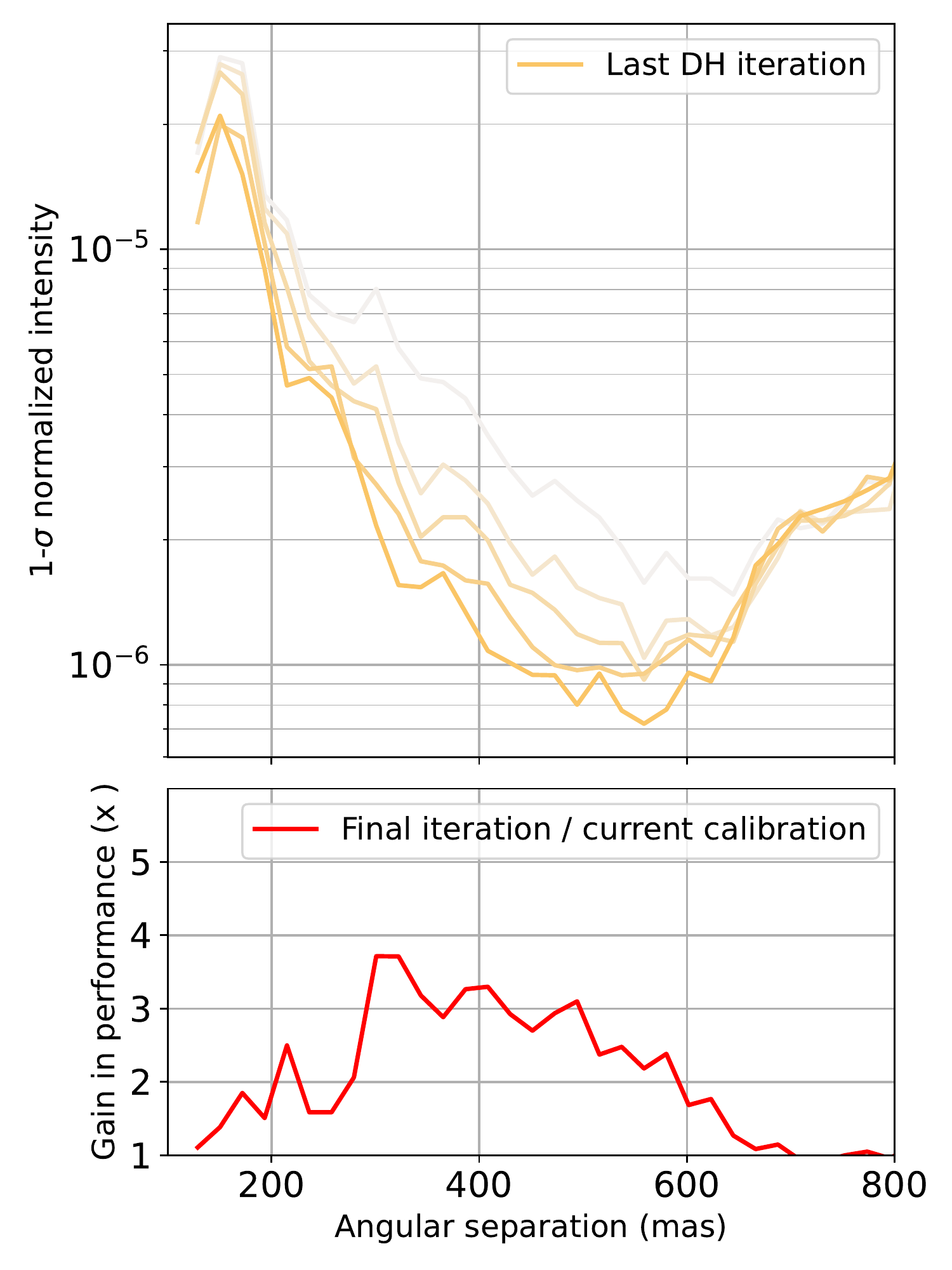}
    \caption{On-sky data. Top: 1-$\sigma$ standard deviation of the normalized intensity obtained at each iteration of the PWP+EFC on-sky pseudo-closed loop in the bottom DH. Bottom: Gain in performance with respect to the current SPHERE calibration (iteration 0).}
    \label{fig:closedloop_onsky}%
\end{figure}
Starting from the original SPHERE calibration, we also closed the correction loop while observing HIP~57013. We applied PWP+EFC in the bottom D-shape DH starting at 135~mas from the star to 650~mas. The exposure time of the probe images was set to 64~s for the sake of averaging the intensity of the turbulence halo below the static speckle intensity. Assuming this halo is stable in time (i.e. the turbulence statistic is in steady state for two probe acquisition in a row), it is then removed in the PWP algorithm when the difference of each pair of diversity images is calculated. In total, each image required 80s with overheads for a total of $\sim$400s per iteration (4 images required for PWP and 1 image to confirm loop convergence). We present the resulting processed images after 4 iterations (i.e. $\sim$30mn) in Fig.~\ref{fig:comparison}. The resulting normalized intensity RMS at each iteration is also plotted in Fig.~\ref{fig:closedloop_onsky}\,, showing that the static speckle intensity is continuously minimized in the DH and demonstrating robustness of the algorithm. A factor $\sim$2 in performance is gained between 165~mas and 270~mas while we achieve a factor $\sim$3 of improvement between 300~mas and 500~mas. Only a few stellar residuals persist in the science image at small separations after 4 iterations. If this signal is coherent with the starlight, we expect these speckles to be corrected with more PWP+EFC iterations or with a better calibration of the PWP diversity amplitude.

\section{Strategy 3: Coherent differential imaging}
\label{sec:cdi}
\begin{figure}[t]
    \centering
    \includegraphics[width=8cm]{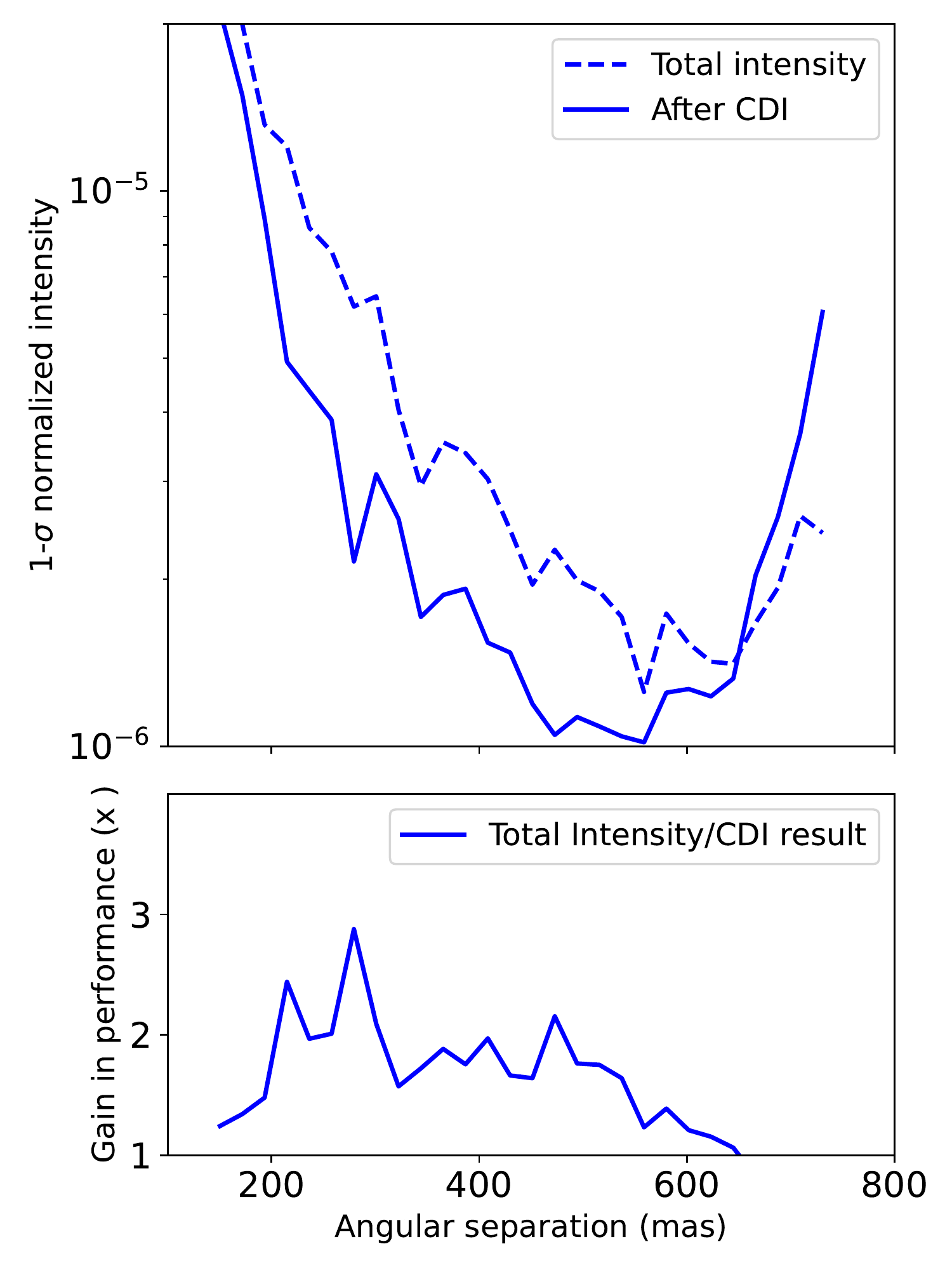}
    \caption{Top: 1-$\sigma$ standard deviation of the normalized high-pass filtered total intensity after one on-sky PWP+EFC iteration (dashed) and processing of the same image with CDI (continuous), calculated in the bottom DH. Bottom: Gain in performance of CDI with respect to the total intensity image at iteration 1.}
    \label{fig:cdi}%
\end{figure}
One can also take advantage of the starlight coherence to extract the incoherent light, including any substellar faint companion, from the total intensity image. Indeed, PWP can be used alone to estimate the stellar speckle E-field. The squared absolute value of this E-field is then used as a reference image to be subtracted from the total intensity image. The same technique has been used to highlight the halo of internal turbulence in Fig.~\ref{fig:internalsouce_contrastinco}. The processed image is then called equivalently unmodulated component, incoherent component, or the result of coherent differential imaging (CDI). When applying PWP+EFC, this post-processing technique can be done for free since PWP is already employed at each iteration. We therefore show the result of CDI after iteration 1 of PWP+EFC in Fig.~\ref{fig:comparison}. This means one single EFC iteration has already been performed in the bottom DH, followed by one PWP set of acquisitions to get the residual speckle E-field that is used for CDI. The resulting CDI image is cleaned out from modt of the stellar speckles in the regions that are well sensed by PWP (top and bottom regions of the field of view for the chosen probes here) and then enable a speckle correction in almost all the field of view. The resulting normalized intensity RMS before and after CDI in the bottom DH is also plotted in Fig.~\ref{fig:cdi}. It demonstrates performance equivalent to the full bottom half DH correction in Sec.~\ref{sec:onsky}, but after less than $\sim15$min of on-sky calibration. The CDI itself improves the performance to a factor up to 3 in the bottom region.

\section{Discussion and conclusion} 
\label{sec:conclusion}
In this work, we have presented three different observing strategies using PWP and EFC. First, minimizing the stellar speckle intensity with a half DH correction performed on the instrument internal calibration unit enables a factor up to 3 in contrast performance. The technique is robust since it is not affected by atmospheric turbulence and it does not overlap with science operations because it uses the AO system during the day. It is also readily adaptable (if anticipated) to any target separation, coronagraph, or spectral bandwidth. However, we showed it has a limited field of view and presented the cause of its performance limitations. The second strategy of observation is to close the loop on-sky while directly observing the target of interest. This method provides the best performance at all angular separation because it is iterative and corrects for wavefront errors along the entire beam, but it relies on good observing conditions and stable AO system and is also limited in field of view. The modulation of the science images with the probes in PWP also prevents a 100\% science duty cycle. The third strategy of observation is to use the PWP estimations to apply CDI in post-processing. This increases the field of view for science and provides encouraging performance but requires PWP to be active during science operations.

In the future, these different strategies can be combined, depending on the science case. First, deeper contrast in one particular region of the image can be achieved using a combination of strategies 1 and 2. Strategy 1 would start improving the instrument performance for free and reduce the required number of iteration for strategy 2, hence increasing the time dedicated to science observations. Strategy 3 can also be applied after strategies 1 and 2 by introducing PWP on the already corrected image to perform a final clean up of the DH and potentially observe fainter objects. It would currently require about 5 min after a few on-sky iterations of PWP+EFC for an improvement factor up to 10 in the half DH region. Strategy 3 could also be applied right after strategy 1 and the speckle calibration would require only 5~min of on-sky observation but with potentially degraded results in the DH region. For a broader field of view and for either the study of extended objects or the blind search for planets, strategy 3 could also be applied alone at a regular cadence while the CDI results would be used an inputs for conventional ADI, SDI, PDI or RDI postprocessing techniques. These different calibration scenarios will be investigated in ongoing studies of the optimal observing strategy for VLT/SPHERE.

\acknowledgments 
 
The research was carried out at the Jet Propulsion Laboratory, California Institute of Technology, under a contract with the National Aeronautics and Space Administration (80NM0018D0004).

\bibliography{bib_GS} 
\bibliographystyle{spiebib} 

\end{document}